\documentclass[conference]{IEEEtran}
\IEEEoverridecommandlockouts 
\usepackage{amsmath,amssymb,amsthm,epsfig,color,empheq,graphicx,graphics}
\usepackage{enumerate,url,epstopdf,enumitem}
\usepackage{xcolor}
\usepackage{float} 
\usepackage{diagbox}

\newtheorem{lemma}{Lemma}
\newtheorem{corollary}{Corollary}

\newtheorem{remark}{Remark}

\newcommand \expect{\mathbb{E}}
\newcommand \sinc {\textrm{sinc}}

\begin{document}

\title{Dynamic Modeling of Load Demand in Electrified Highways Based on the EV Composition}

\author{
    \IEEEauthorblockN{Ashutossh Gupta, Vassilis Kekatos, Dionysios Aliprantis, and Steve Pekarek }
    \IEEEauthorblockA{\textit{Elmore Family School of Electrical and Computer Engineering}, \textit{Purdue University}, West Lafayette, IN, USA\\
    \texttt{\{gupta799,kekatos,dionysios,spekarek\}@purdue.edu}
    }
    \thanks{This work was supported in part by US NSF grants 2500682 and 1941524.}
}

\maketitle
\vspace*{-3em}
\begin{abstract}
Electrified roadways (ERs) equipped with the dynamic wireless power transfer (DWPT) technology can achieve longer driving range and reduce on-board battery requirements for electric vehicles (EVs). Due to the spatial arrangement of transmitter (Tx) coils embedded into the ER pavement, the power drawn by the EV's receiver (Rx) coil is oscillatory in nature. Therefore, understanding the dynamic behavior of the total DWPT load is important for power system dynamic studies. To this end, we model the load of individual EVs in the time and frequency domains for constant EV speed. We establish that a nonlinear control scheme implemented in existing DWPT-enabled EVs exhibits milder frequency harmonics compared to its linear alternative. According to this model, the harmonics of an EV load decreases in amplitude with the Rx coil length. We further propose and analyze stochastic models for the total DWPT load served by an ER segment. Our models explains how the EV composition on the ER affects its frequency spectrum. Interestingly, we show that serving more EVs with longer Rx coils (trucks) does not necessarily entail milder harmonics. Our analytical findings are corroborated using realistic flows from a traffic simulator and offer valuable insights to grid operators and ER designers.
\end{abstract}

\begin{IEEEkeywords}
 EV charging, oscillations, power spectrum.
\end{IEEEkeywords}

\section{Introduction}
\allowdisplaybreaks
Charging EVs wirelessly and during transit has been proposed as a solution to eliminate range anxiety and reduce onboard battery capacity requirements~\cite{Diala19}. Electrified roadways (ER) are being developed with transmitter coils embedded within them. When an EV instrumented with a receiver (Rx) coil moves over the ER, it can absorb power through electromagnetic induction. The power gained from the ER can be used to propel EVs or also charge its battery~\cite{vatan2025receiver}. The Tx coils present in an ER segment are powered by a substation. The total load at the substation is the sum of the power drawn by all EVs served by the ER. As an EV moves over the ER, the spacing between the Tx coils causes the DWPT load to exhibit oscillatory behavior~\cite {Diala19}. This work aims to study the effect of the EV composition on the frequency spectrum of the DWPT load served by a substation.

Prior works exploring the impact of DWPT systems on power grids focus primarily on the average (DC component) consumed by the ER~\cite{sauter2024power}. Reference~\cite{newbolt2023diverse} proposes a battery operation strategy to reduce voltage deviations across the distribution grid powering the ER, while~\cite{ghose2025trafficawaregridplanning} optimally sizes a microgrid to support the ER infrastructure. 

However, time-varying loads may introduce frequency harmonics that can trigger power system oscillations~\cite{chen2022fast},~\cite{Alizadeh2025Ontario}. If these harmonics occur at frequencies below $2$~Hz, such loads might pose the risk of inter-area oscillations~\cite{Sun2023realtime}. Therefore, it becomes pertinent to study the dynamic (frequency-domain) characteristics of the total DWPT load consumed by an ER. In~\cite{gupta2025frequency}, we studied the location and magnitude of harmonics introduced by DWPT loads under diverse traffic scenarios. Nonetheless, the analysis presumed EV loads were scaled versions of a reference load waveform and ignored the effect of different Rx coil lengths and peak demands across EVs.

To better reflect the heterogeneity of EV populations on dynamic DWPT models, this work contributes on four fronts: \emph{c1)} We devise a DWPT load model to capture the nonlinear control scheme of the EV controllers (Sec.~\ref{sec:1tdmodel}); \emph{c2)} For a single EV load the first harmonic dominates the rest and that EVs with longer Rx coils feature weaker harmonics (Sec.~\ref{sec:1fdmodel}); \emph{c3)} We evaluate the spectrum of the total DWPT load when the ER serves EVs of different classes (Sec.~\ref{sec:total}); and \emph{c4)} Our findings invalidate the conjecture that if an ER serves more trucks than sedans, harmonics in the total DWPT load are weaker (Sec.~\ref{sec:EV composition}). Numerical tests using a traffic simulator corroborate our findings in Sec.~\ref{sec:numerical_tests}.

\begin{figure}[t]
\centering
\vspace{-1em}
\includegraphics[width=\columnwidth]{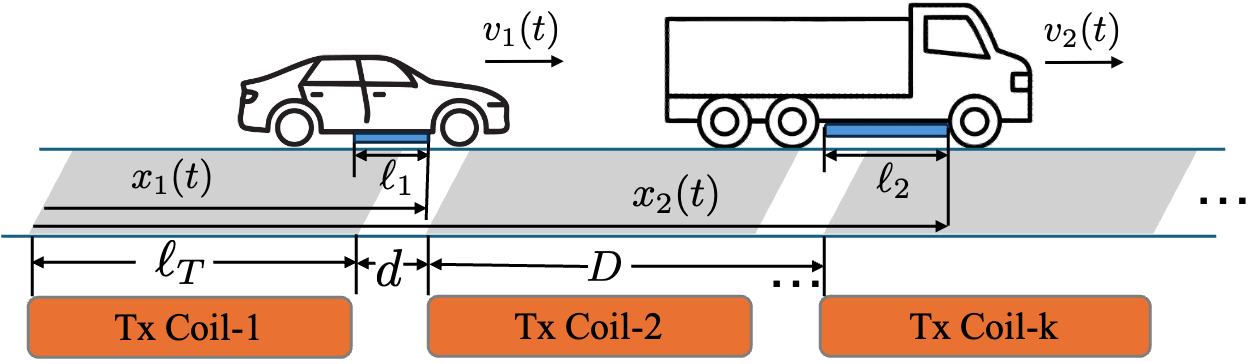}
\vspace{-1.5em}
\caption{In a DWPT-enabled ER, transmitters Tx of length $\ell_T$ and gaps of length $d$ are arranged periodically, every $D = \ell_T + d$~meters. An EV draws power when its Rx overlaps with Tx coils. Because of their higher power requirements, trucks have longer Rx lengths than sedans ($\ell_2 > \ell_1$).}
\vspace{-1.5em}
\label{fig:ER}
\end{figure}

\section{Time-Domain Modeling of a Single DWPT Load}\label{sec:1tdmodel}
We consider traffic flow in a single direction and assume the DWPT technology has been deployed on a single lane of the ER~\cite{bafandkar2025charging}. The goal is to model the total DWPT load consumed at a substation energizing an ER segment of length $L$ in meters. This section reviews a time-domain model of a single EV load. The ER powers EVs through electromagnetic induction. The ER is instrumented with transmitter (Tx) coils of length $\ell_T$ meters laid under the pavement. To meet road construction standards, every Tx coil is followed by a $d$-meter-long gap; see Fig.~\ref{fig:ER}. A Tx coil and the subsequent gap form a \emph{coil segment} of length $D = \ell_T +d$. Coil segments are indexed by $k=0,\ldots,K-1$, where $K = L/D$. 

Suppose this ER segment powers $N$ EVs, indexed by $n=1,\ldots,N$. Each EV is instrumented with a receiver (Rx) coil. Rx$_n$ will denote the Rx coil of EV$_n$ and is of length $\ell_n$ in meters. EVs with higher power requirements (electric trucks) may have larger $\ell_n$. We assume $\ell_T>\ell_n\geq d $ for all $n$. Let $x_n$ denote the position of Rx$_n$'s front end with respect to the beginning of the ER segment; see Fig.~\ref{fig:ER}. 

\begin{figure}[t]
\centering
\includegraphics[width=0.87\linewidth]{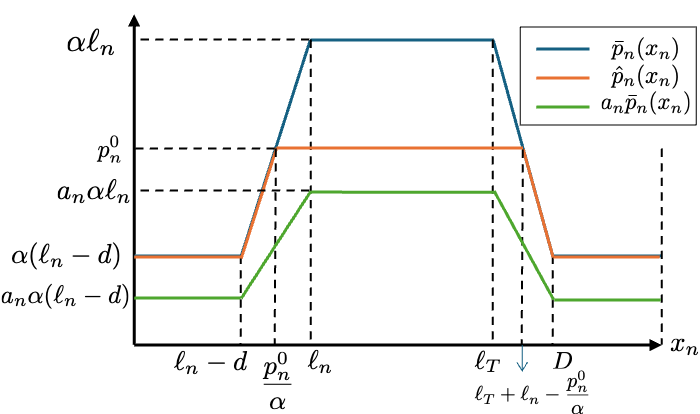}
\vspace{-1em}
\caption{The blue curve depicts the maximum DWPT load $\bar{p}_n(x_n)$ as a function of EV position $x_n$. The orange curve depicts the clipped load $\hat{p}_n(x_n)$ for peak power demand $p_n^d \in(\alpha (\ell_n-d), \alpha \ell_n ]$. The green curve depicts $\bar{p}_n(x_n)$ scaled by a factor of $a_n\in[0,1]$; see~\cite{gupta2025frequency}.} 
\label{fig:trapezoid}
\vspace{-1.75em}
\end{figure}

The maximum power that EV$_n$ can consume while being at position $x_n$ is determined by \emph{i)} the overlap between Rx$_n$ and the Tx coils; and \emph{ii)} the \emph{rated spatial power density} $\alpha$ the ER can supply, measured in kW/m of Rx/Tx overlap~\cite{gupta2025frequency}. Due to the repetitive layout of Tx coils, the maximum power $\bar{p}_n(x_n)$ EV$_n$ can consume when located at position $x_n$ is a periodic trapezoidal waveform. The spatial period of this waveform is $D$ and the repeating trapezoidal pulse is shown in blue in Fig.~\ref{fig:trapezoid}. Because the Rx/Tx overlap lies in the range $[\ell_n-d,\ell_n]$, the waveform $\bar{p}_n(x_n)$ ranges within $(\alpha (\ell_n-d),\alpha\ell_n]$.

Depending on its power needs, the energy management system of EV$_n$ may decide to consume less power than $\bar{p}_n(x_n)$. Let $\hat{p}_n(x_n)$ denote the power actually consumed by EV$_n$ at position $x_n$, with $\hat{p}_n(x_n)\leq \bar{p}_n(x_n)$. In current DWPT implementations, the EV energy management system specifies the peak demand $p_n^d$. A converter saturates the consumed power to $p_n^d$, so that $\hat{p}_n(x_n)\leq p_n^d$ for all $x_n$~\cite{vatan2025receiver}. We termed this method of DWPT load control as the \emph{clipping control scheme}~\cite{gupta2025frequency}. The consumed power can be expressed as
\begin{equation} \label{eq:power(x)}
\hat{p}_n(x_n) = 
\begin{cases}
p_n^d, &  \bar{p}_n(x_n)\geq p_n^d\\
\bar{p}_n(x_n), & \textrm{otherwise}.
\end{cases}
\end{equation}
The DWPT load of EV$_n$ over a single coil segment takes the trapezoidal form shown as the orange waveform in Fig.~\ref{fig:trapezoid}:
\begin{equation*} 
    g_n(x_n) = 
    \begin{cases}
        \alpha (\ell_n - d),  & 0 \leq x_n < \ell_n-d \\
        \alpha x_n, & \ell_n-d \leq x_n < \tfrac{p_n^d}{\alpha} \\
        p_n^d, & \tfrac{p_n^d}{\alpha}   \leq x_n < \ell_n+\ell_T-\tfrac{p_n^d}{\alpha}\\
        \alpha (\ell_n + \ell_T - x_n), & \ell_n+\ell_T-\tfrac{p_n^d}{\alpha} \leq  x_n < D.
    \end{cases}
\end{equation*}
Across all coil segments, the load $\hat{p}_n(x_n)$ is a finite-duration, periodic function with period $D$:
\begin{equation} \label{eq:pn(t)}
\hat{p}_n(x_n) = \sum_{k=0}^{K-1} g_n(x_n-kD).   
\end{equation}
This holds if $p_n^d\in (\alpha (\ell_n-d),\alpha\ell_n]$. For $p_n^d=\alpha\ell_n$ in particular, the orange waveform coincides with the blue one. On the other hand, if $p_n^d\in [0,\alpha (\ell_n-d)]$, the orange waveform becomes constant and $\hat{p}_n(x_n)=p_n^d$ for all $x_n$. 

To analyze the DWPT load in the frequency domain, we must express it as a function of time $t$ rather than position $x_n$. For simplicity, suppose that all EVs move at constant speed $v$, measured in m/s. Our analysis could be generalized to the setting of different EV speeds, similarly to~\cite{gupta2025frequency}. Let $t_n$ denote the time in seconds when EV$_n$ enters the ER segment. Hence, the position of EV$_n$ can be expressed as
\begin{equation}\label{eq:xnt}
x_n(t) = v \cdot (t-t_n)\cdot u(t-t_n),
\end{equation}
where $u(t)$ is the unit step function. If we substitute \eqref{eq:xnt}, we can express the DWPT load as a periodic function of time 
\begin{equation} \label{eq:power(t)}
p_n(t) = \sum_{k=0}^{K-1} g_n\left(v(t-t_n-kT)\right)u(t-t_n)
\end{equation}
with fundamental period $T=D/v$~s. If $p_n^d \leq \alpha (\ell_n-d)$, the DWPT load $p_n(t)$ remains constant at $p_n^d$ at all times.

\emph{\textbf{INDOT ER testbed.}} Before studying the DWPT spectrum, let us summarize the technical characteristics of a pilot ER project~\cite{vatan2025receiver}. In this pilot, $\ell_T = 3.66$~m, $d=0.91$~m-long gap, $D=4.57$~m, and $\alpha=109.36$~kW/m. The speed limit is $v=24.6$~m/s. This testbed includes a prototype DWPT-enabled Class-8 truck with $\ell_n=1.83$~m.



\section{Frequency Spectrum of Single DWPT Load} \label{sec:1fdmodel}
This section develops a frequency-domain model for a single DWPT load $p_n(t)$. We consider the case where $p_n^d \in (\alpha (\ell_n-d),\alpha \ell_n]$, so that $p_n(t)$ is periodic. The load $p_n(t)$ is a train of $K$ trapezoidal pulses per \eqref{eq:pn(t)}. Because the number of coil segments $K$ is large, the EV load can be approximated by an infinite-duration periodic waveform of trapezoidal pulses. Suppose for now that $t_n=0$. Then, the DWPT load $p_n(t)$ can be approximated by the Fourier series (FS) expansion
\begin{equation}\label{eq:gn}
p_n(t)\simeq  \sum_{m=-\infty}^{\infty} c_{m,n} e^{jm\omega_0t},
\end{equation}
where $\omega_0 = \frac{2\pi v}{D}$ is the fundamental frequency and $c_{m,n}$ is the $m$-th FS coefficient. We shift $g_n(t)$ so it is even-symmetric around $t=0$ and the FS coefficients $c_{m,n}$ are real-valued. 

The FS coefficient of the DC term in \eqref{eq:gn} is the power actually consumed by EV$_n$ and can be found to be
\begin{equation}\label{eq:c0n}
c_{0,n}= \frac{p_n^d}{D} \left(\ell_T + \ell_n - \frac{p_n^d}{\alpha}\right).
\end{equation}

At maximum peak demand $p_n^d=\alpha \ell_n$, the DC term becomes $c_{0,n}=\alpha \ell_n\frac{\ell_T}{D}$. In this latter case, the power consumed by EV$_n$ scales with its Rx length $\ell_n$, the power density $\alpha$ of the ER, and the duty cycle $\frac{\ell_T}{D}$ of the ER. The FS coefficient for the $m$-th harmonic can be found to be for all $m\geq 1$
\begin{equation}\label{eq:cmn}
c_{m,n}=c_{0,n}\sinc \left( \frac{m p_n^d}{\alpha D} \right) \sinc \left(\frac{m}{D} \left(\ell_T + \ell_n - \frac{p_n^d}{\alpha}\right)\right),
\end{equation}
where $\sinc(x) := \frac{\sin(\pi x)}{\pi x}$ is the normalized $\sinc$ function. 

For any arbitrary timing $t_n\neq 0$, the FS expansion of \eqref{eq:gn} is adjusted per the time-shifting property of the FS as
\begin{equation} \label{eq:FS of pn(t)}
    p_n(t) = \sum_{m=-\infty}^{\infty} c_{m,n} e^{-jm\omega_0 t_n} e^{jm\omega_0 t}.
\end{equation}

To capture the relative harmonic content of $p_n(t)$ over its DC term, we define the metric of \emph{total harmonic content}:
\begin{equation}\label{eq:THC_n}
\text{THC}_n = \sqrt{2 \sum_{m=1}^{\infty}\left(\frac{c_{m,n}}{c_0,n}\right)^2} \times 100\%.
\end{equation}
In the INDOT ER if $\ell_n = 1.83~$m and $p_n^d = \alpha \ell_n = 200~$kW, we obtain THC$_n = 26 \%$. Three results are in order. 

\begin{remark}\label{re:firstharmonic}
Because $|\sinc(x)|\leq |\pi x|^{-1}$, the FS coefficients for $m\geq 1$ can be upper bounded as
\begin{equation}\label{eq:cm bound}
|c_{m,n}| \leq c_{0,n} \frac{\alpha D^2}{m^2 \pi^2 p_n^d \left(\ell_T + \ell_n -\frac{p_n^d}{\alpha}\right)}= \frac{\alpha D}{m^2 \pi^2}.
\end{equation}
Given that $|c_{m,n}|$ decreases quadratically in $m$, the first harmonic is much stronger than the rest. For instance, in the INDOT ER example, the first harmonic exhibits a relative signal power of $25\%$ with respect to the DC term. 
\end{remark}

\begin{remark}\label{re:elln}
Given Remark~\ref{re:firstharmonic}, to explore the effect of Rx length on the THC, it suffices to study how the ratio $c_{1,n}/c_{0,n}$ depends on $\ell_n$. If EV$_n$ consumes maximum power $p_n^d = \alpha \ell_n$, we get
\begin{equation} \label{eq:c1/c0 }
    \frac{c_{1,n}}{c_{0,n}} = \sinc \left( \frac{\ell_n}{D} \right) \sinc \left( \frac{\ell_T}{D} \right).
\end{equation}
Since $0<\ell_n<\ell_T<D$, the ratio decreases monotonically with $\ell_n$. Hence, the DWPT loads of EVs with longer Rx coils (trucks) have weaker harmonics relative to their DC power.
\end{remark} 

Different from the clipping control scheme of \eqref{eq:power(x)}, in~\cite{gupta2025frequency} we considered a \emph{scaling control scheme}, according to which the EV load is a scaled down version of $\bar{p}_n(x_n)$ so that
\begin{equation}\label{eq:scaling}
p_n^s(t) = a_n \bar{p}_n(x_n(t)).
\end{equation}
where $a_n\in(0,1]$. The scaled load is shown in green in Fig.~\ref{fig:trapezoid}. Lemma~\ref{le:control} compares the two schemes in terms of $c_{1,n}/c_{0,n}$.

\begin{lemma}\label{le:control}
If $\ell_T > D/2$, the clipping control scheme of \eqref{eq:power(x)} attains a smaller harmonic ratio $c_{1,n}/c_{0,n}$ than the scaling control scheme of \eqref{eq:scaling} for all $(p_n^d,\ell_n)$.
\end{lemma}



\section{Frequency-Domain Model of Total DWPT Load}\label{sec:total}
Upon ignoring losses, the total DWPT load is the sum of EV loads. If all EVs move at the same constant speed $v$, the load $p(t)$ is periodic and is amenable to an FS expansion:
\begin{subequations}
\begin{align}
 p(t) &= \sum_{n=1}^N p_n(t)=\sum_{m=-\infty}^{\infty} c_m e^{jm\omega_0t}, \label{eq:total_power}\\
\textrm{where}\quad c_m &= \sum_{n=1}^N c_{m,n} e^{-jm\omega_o t_n}, \label{eq:cm}
\end{align}     
\end{subequations}
where \eqref{eq:cm} is obtained by plugging \eqref{eq:FS of pn(t)} into \eqref{eq:total_power}. Note that EV parameters $(\ell_n,p_n^d)$ affect only $c_{m,n}$, whereas the EV timing $t_n$ affects only $e^{-jm\omega_o t_n}$ in \eqref{eq:cm}.  Given the uncertainty in $\{\ell_n,p_n^d,t_n\}_{n=1}^N$, we model them as random variables. Hence, the FS coefficients $c_m$ are random and $p(t)$ is a stochastic signal. Because $(\ell_n,p_n^d)$ are independent of $t_n$, random variables $c_{m,n}$ and $t_n$ are independent. It is practical to model $c_{m,n}$'s as independent across EVs. Ignoring any correlation across EV timings, we model $t_n$'s as independent and uniformly drawn from $[0,T)$.


To study the frequency content of $p(t)$, one needs to evaluate its \emph{power spectral density} (PSD)~\cite[Ch.~4]{proakis}. The PSD shows how the signal power is distributed across frequencies. A sufficient condition for the PSD of a stochastic signal to exist is that the signal is \emph{wide sense stationary (WSS)}, i.e., its mean and autocorrelation functions are time-invariant. The ensuing lemma establishes that $p(t)$ is WSS and finds its PSD. 

\begin{lemma}\label{le:stationary}
If EV timings $t_n$ are independently and uniformly within $[0,T)$, the total DWPT load $p(t)$ in \eqref{eq:total_power} is WSS, its mean is $\mu_p(t)= N \expect[c_{0,n}]$, and its PSD is
\begin{subequations}\label{eq:lemma1}
\begin{align}
 S_p(\omega) = 2 \pi &\sum_{m=-\infty}^{\infty} \expect [|c_m|^2] \cdot \delta(\omega - m\omega_0),\label{eq:lemma1:a}\\
\text{where}~~~\expect \left[|c_m|^2\right] &=
\begin{cases}
     N^2 \bigl(\expect[c_{0,n}] \bigr)^2, \quad &m = 0 \\
    N \expect [c_{m,n}^2], & m \neq 0.
\end{cases}\label{eq:lemma1:b}
\end{align}     
\end{subequations}
The expectation in \eqref{eq:lemma1:b} applies over the distribution of $c_{m,n}$.
\end{lemma}


According to Lemma~\ref{le:stationary}, the PSD of $p(t)$ consists of discrete components at the harmonic frequencies having magnitude $\expect[|c_m|^2]$. If the ER powers EVs moving at multiple different speeds, the PSD will contain harmonic sets associated with each speed. Analogous to the THC of the single EV load, we can define the THC of the total DWPT load $p(t)$ as
\begin{equation}\label{eq:THC}
\text{THC}_p= \sqrt{\frac{2\sum_{m=1}^{\infty} \expect [|c_m|^2]}{\expect [|c_0|^2]}} =\sqrt{\frac{2\sum_{m=1}^{\infty} \expect [|c_{m,n}|^2]}{N\left(\expect [c_{0,n}]\right)^2}}
\end{equation}
where the second equality follows from \eqref{eq:lemma1:b}. If squared, the metric THC$_p$ is the ratio of the total power of $p(t)$ distributed across harmonics over the power of its DC component. Next, we study how THC$_p$ varies under different EV populations.


\section{Effect of EV Composition on DWPT Spectrum}\label{sec:EV composition}
This section studies how the composition of the EV population served affects the THC of the total DWPT load. We partition EVs into $G$ \emph{classes} with the same Rx length. For class $g=1,\ldots,G$, let $\ell_g$ denote the Rx length and $\pi_g$ the probability of an EV belonging to this class. Therefore, the random variable $\ell_n$ is drawn from a probability mass function (PMF) taking $G$ discrete values $\{\ell_g\}_{g=1}^G$, occurring with probability $\{ \pi_g\}_{g=1}^G$. If $N_g$ is the number of EVs of class $g$, it follows that $\pi_g= N_g/N$. Peak power demands $p_n^d$ can be drawn from a continuous probability distribution dependent on $\ell_n$, to capture heterogeneity in EV loads within the same class. Under this model, the expectations needed to compute the PSD from \eqref{eq:lemma1}--\eqref{eq:THC} can be expressed as
    \begin{subequations}\label{eq:cond}
        \begin{align}
            \expect[c_{0,n}] &= \sum_{g=1}^G \pi_g \, \expect[c_{0,n} \mid \ell_n = \ell_g]\label{eq:cond:0} \\
            \expect[c_{m,n}^2] &= \sum_{g=1}^G \pi_g \, \expect[c_{m,n}^2 \mid \ell_n = \ell_g]. \label{eq:cond:m}
        \end{align}
    \end{subequations}
For instance, peak demands $p_n^d$ can be drawn independently uniformly in $(\alpha (\ell_g-d), \alpha \ell_g]$. In this case, the final expression for \eqref{eq:cond} reveal that although $\expect [c_{0,n}^2]$ increases with $\ell_g$'s, the dependence of $\expect [c_{m,n}^2]$ on $\ell_g$'s is less obvious.

To better understand how THC$_p$ depends on the EV composition, we simplify the analysis and assume maximum peak demand $p_n^d = \alpha \ell_n$. Then, all EVs in class $g$ share the same FS coefficients $c_{m,n} = c_m^g$ for all $n$ of class $g$. This implies that $c_{m,n}$ is a random variable drawn from a PMF taking  values $\{c_m^g\}_{g=1}^G$ with probability $\{ \pi_g \}_{g=1}^G$. In this case, we obtain
\begin{subequations}\label{eq:moments_simplified}
\begin{align}
&\expect [ c_{0,n}]=\sum_{g=1}^G \pi_g c_m^g = \sum_{g=1}^G \alpha\pi_g \ell_g\frac{\ell_T}{D}   \label{eq:mean_simplified} \\
&\expect [ c_{m,n}^2]{=}\sum_{g=1}^G \pi_g\left[\frac{\alpha D}{m^2 \pi} \sin \left( \frac{m\pi \ell_g}{D} \right) \sin \left(\frac{m \pi \ell_T}{D} \right)\right]^2 \label{eq:moment_simplified}
\end{align}    
\end{subequations}
Plugging \eqref{eq:moments_simplified} into \eqref{eq:THC} provides the THC$_p$. Observe that the DC term increases if EV classes with longer Rx lengths are more populous. Because $\expect [c_{m,n}^2]$ decays with $m^4$, we approximate THC$_p$ by considering only the first harmonic. 

Per Remark~\ref{re:elln}, the THC of a single DWPT load decreases with $\ell_n$. One may conjecture that the THC of the total DWPT load decreases when EV classes featuring longer $\ell_n$ are more populous. Corollary~\ref{co:EV composition} shows the conjecture is not always true.

\begin{corollary}\label{co:EV composition}
Suppose the ER serves only two EV classes having Rx lengths $\ell_a$ (trucks) and $\ell_b$ (sedans) with $0<\ell_b<\ell_a<D$. Consider two traffic scenarios S1) and S2) having a total number of EVs $N_1$ and $N_2$, respectively. Let $\theta_1$ and $\theta_2$ denote the fraction of trucks per scenario with $\theta_1 > \theta_2$. Select parameters $(N_1, \theta_1, N_2, \theta_2)$ so that the DC component of the total DWPT load is identical for S1) and S2). Then, the \textrm{THC}$_p$ under S1) is less than the \textrm{THC}$_p$ under S2) if and only if 
\begin{equation} \label{eq:condition}
\ell_a \sin^2 \left( \frac{\pi \ell_b}{D} \right)  > \ell_b \sin^2 \left( \frac{\pi \ell_a}{D} \right).
\end{equation}
\end{corollary}

Corollary~\ref{co:EV composition} predicates that serving more trucks than sedans does not necessarily entail a lower THC. For the INDOT ER, if trucks are instrumented with Rx of lengths $\ell_a = 1.83$~m, the inequality above holds for $\ell_b \in (1.51, 1.83)$~m. If $\ell_b<1.51$~m, higher penetration of trucks incurs higher THC. Corollary~\ref{co:EV composition} fixed the DC terms and let the $(N_1,N_2)$ vary. Interestingly, condition \eqref{eq:condition} compares the two THCs even if we set $N_1=N_2$ and let the DC terms vary. The following numerical tests validate \eqref{eq:condition} under more general settings.

\section{Numerical Tests and Conclusions} \label{sec:numerical_tests}


We simulated realistic vehicle trajectories using SUMO, an open-source microscopic traffic flow simulator~\cite{SUMO18}. After simulating the traffic flow on the INDOT ER of length $L=4$~km, DWPT loads were obtained using~\eqref{eq:power(x)} and~\eqref{eq:total_power}. Two vehicle classes (trucks and sedans) were considered. Trucks and sedans were assigned maximum speeds of $22$ and $30$~m/s, respectively. The mean EV arrival rate was $0.21$~EV/s. 

\begin{figure}[t]
\centering
\includegraphics[width=0.8\linewidth]{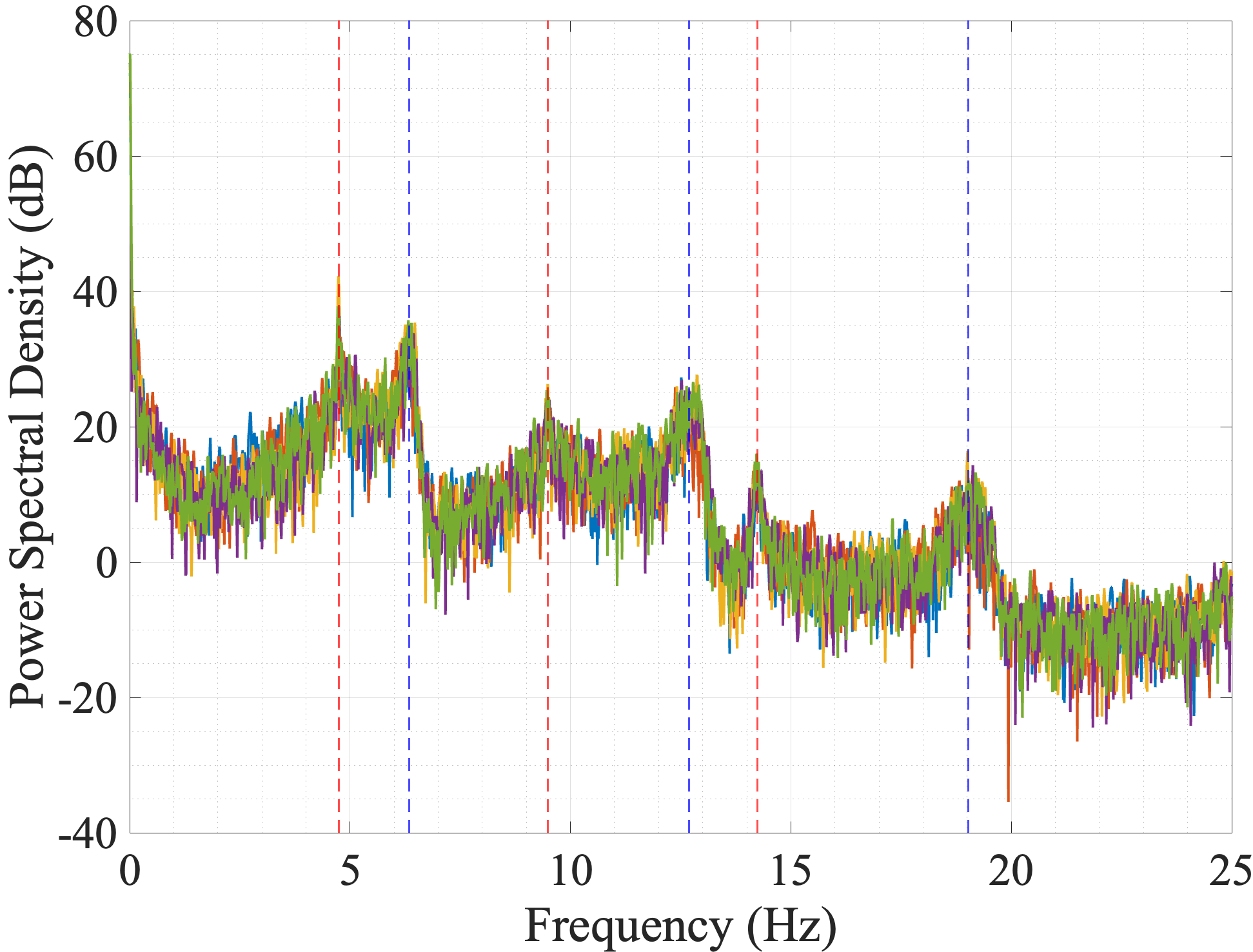}
\vspace{-1em}
\caption{Numerically computed PSD of $1$-min DWPT loads. We observe two sets of harmonics of $4.74~$Hz (red) and $6.34$~Hz (blue).}
\label{fig:PSD}
\vspace{-1em}
\end{figure}

In the first test, we wanted to test whether Lemma~\ref{le:stationary} explains the DWPT spectrum. We simulated trucks with $\ell_a = 1.83$~m and sedans with $\ell_b = 1.2$~m. Peak power demands $p_n^d$ were drawn uniformly from $[50,200]$~kW. Figure~\ref{fig:PSD} shows the numerically computed PSD of 1-min-long DWPT loads. As trucks and sedans have different maximum speeds, the spectrum contains two sets of harmonics corresponding to speeds $21.7$ and $29$~m/s. Note that the PSD spreads around the harmonic peaks due to the finite signal length; see~\cite{gupta2025frequency}. Overall the key features of spectra in Fig.~\ref{fig:PSD} agrees with Lemma~\ref{le:stationary}.

In the second test, we examined the validity of Corollary~\ref{co:EV composition}. To this end, we compared the average THC of multiple $1$-minute DWPT loads, obtained at different truck penetrations $\theta$ and sedan Rx lengths $\ell_b$. The total number of EVs was fixed to $N = 45$ for all experiments. We observed that when $\ell_b$ takes values $0.58$ and $1.2$~m, the THC$_p$ increases with $\theta$. This is expected as the \eqref{eq:condition} is not satisfied for $\ell_b < 1.51$~m. We also observed THC remains almost constant across varying truck penetrations for $\ell_b = 1.7$~m, because for $\ell_b \in (1.51, 1.8)$~m harmonic ratios are close to $1$. 


\begin{table}[t]
\centering
\caption{Average THC (\%) for Different Truck Penetrations $(\theta$)\\ and Sedan Rx Lengths $(\ell_b)$}
\vspace{-1em}
\begin{tabular}{|r|r|r|r|}
\hline
\diagbox{$\theta(\%)$}{$\ell_b$ (m)}
    & 0.58
    & 1.20
    & 1.70  \\
\hline
3.77 & \multicolumn{1}{r|}{8.05} & \multicolumn{1}{r|}{7.02} & \multicolumn{1}{r|}{6.65}  \\
\hline
5.57 & \multicolumn{1}{r|}{9.37} & \multicolumn{1}{r|}{7.31} & \multicolumn{1}{r|}{6.67}  \\
\hline
17.75 & \multicolumn{1}{r|}{10.73} & \multicolumn{1}{r|}{8.88} & \multicolumn{1}{r|}{6.82} \\
\hline
\end{tabular}
\label{tab:THC}
\vspace{-2em}
\end{table}

In conclusion are approximate models accurately predict location and magnitude of harmonics of realistic DWPT loads. The condition on Rx lengths does hold in practice as the first harmonic is of highest magnitude. This work can be extended in several directions, such as sizing and designing BESS systems to smooth out the DWPT load and/or analyzing the effect of such loads on power grid frequency dynamics.



 \color{black}

\section*{Appendix}\label{sec:appendix}

\begin{IEEEproof}[Proof of Lemma~\ref{le:control}] From~\eqref{eq:cmn}, the ratio $c_{1,n}/c_{0,n}$ under the clipping control scheme is
\begin{equation}
    f(p_n^d) = \sinc\left( \frac{p_n^d}{\alpha D}\right) \sinc \left( \frac{1}{D} \left(\ell_T + \ell_n - \frac{p_n^d}{\alpha} \right) \right). 
\end{equation}
The ratio under the scaling scheme is $f(\alpha \ell_n)$ for all $p_n^d$. The derivative $f'(p_n^d)$ is positive for $p_n^d \in (\alpha (\ell_n-d), \alpha \ell_n]$ if
\[ \sinc \left( \frac{1}{D} \left(\ell_T + \ell_n - \frac{p_n^d}{\alpha} \right) \right) > \cos \left( \frac{\pi}{D} \left( \ell_T + \ell_n - \frac{p_n^d}{\alpha}  \right) \right). \]
This condition is satisfied for all $p_n^d \in (\alpha (\ell_n-d), \alpha \ell_n]$ if $\ell_T > D/2$. Then, $f(p_n^d)$ is an increasing function, and so $f(p_n^d) \leq f(\alpha \ell_n)$. For $p_n^d \leq \alpha (\ell_n-d)$, clipped DWPT loads become constant $(c_{1,n}=0)$, whereas scaled loads exhibit the same harmonic ratio of $f(\alpha\ell_n)$.\end{IEEEproof}
\begin{IEEEproof}[Proof of Lemma~\ref{le:stationary}] From \eqref{eq:total_power}, the mean of $p(t)$ is
\begin{equation}\label{eq:E[p(t)]}
    \expect [p(t)] = \sum_{m=-\infty}^{\infty} \expect [c_m] e^{jm\omega_0t}.
\end{equation}
To find $\expect [c_m]$, apply the expectation operator on~\eqref{eq:cm} to get
\begin{equation}\label{eq:E[cm_tilde]}
\expect [c_m]= \sum_{n=1}^N \expect[c_{m,n}] \cdot \expect[e^{-jm\omega_0t_n}]= N \expect[c_{m,n}]\cdot \delta[m], 
\end{equation}
where $\delta[m]$ is the Kronecker delta function. The second equality holds because $t_n$ is drawn uniformly in $[0,T]$ so that $\expect [e^{-jm\omega_0 t_n}] = \delta[m]$.
Substituting~\eqref{eq:E[cm_tilde]} in~\eqref{eq:E[p(t)]} provides $\mu_p(t)= N \expect[c_{0,n}]$ or that the mean of $p(t)$ is time-invariant. The autocorrelation $R_p(t,t+\tau)=\expect \left[p(t) p^*(t+\tau)\right]$ is
\begin{equation*}\label{eq:Ra,Rc}
R_p(t,t+\tau)=\sum_{m=-\infty}^{\infty}\sum_{k=-\infty}^{\infty} \expect [c_mc_k^*]  e^{jm\omega_0t}  e^{-jk\omega_0(t+\tau)}.
\end{equation*}
Among all $\expect [c_mc_k^*]$ terms, those with $k \neq m$ are zero because
\begin{equation*}
\expect [c_mc_k^*]=\expect [c_m]\cdot  \expect [c_k^*] 
=\expect[c_{m,n}] \cdot \expect[c_{k,n}^*] \cdot \delta[m] \cdot \delta[k]=0.
\end{equation*}
Hence, the autocorrelation function of $p(t)$ is time-invariant
\begin{equation}\label{eq:Ra_tau}
R_p(t,t+\tau)=R_p(\tau)=\sum_{m=-\infty}^{\infty} \expect [|c_m|^2] \cdot e^{-jm\omega_0 \tau},
\end{equation}
where $\expect [|c_m|^2]$ can be computed from \eqref{eq:cm} as
\begin{equation*}
\expect \left[|c_m|^2\right] =\sum_{n=1}^N \sum_{\ell=1}^N \expect \left[c_{m,n}c_{m,\ell}\right] \expect \left[e^{-jmw_0 t_n} e^{jmw_0t_\ell}\right].
\end{equation*}
For $m\neq0$, only the terms with $n=l$ remain. For $m=0$, all terms contribute to the DC component. Since $p(t)$ is WSS, its PSD is found as the Fourier transform of $R_p(\tau)$~\cite{proakis}.
\end{IEEEproof}

\color{blue}

\color{black}



\begin{IEEEproof}[Proof of Corollary~\ref{co:EV composition}]
Scenario \emph{S1)} has $\theta_1N_1$ trucks and $(1-\theta_1)N_1$ sedans, while \emph{S1)} has $\theta_2N_2$ trucks and $(1-\theta_2)N_2$ sedans. Equating the DC terms under \emph{S1)} and \emph{S2)} yields
\begin{equation*}
c_0^a \theta_1 N_1 + c_0^b (1-\theta_1) N_1=
c_0^a \theta_2 N_2 + c_0^b (1-\theta_2) N_2. 
\end{equation*}
Therefore, the ratio of EV numbers of the two scenarios is
\begin{equation}\label{eq:N1/N2}
    \frac{N_1}{N_2} = \frac{c_0^a \theta_2 + c_0^b (1-\theta_2) }{c_0^a \theta_1 + c_0^b (1-\theta_1)}.
\end{equation}

Granted that the first harmonic dominates over the others, it suffices to compare $\expect[|c_1|^2]$ for the two traffic scenarios. The ratio of $\expect[|c_1|^2]$ under \emph{S1)} and \emph{S2)} can be expressed as
\begin{equation}\label{eq:c1_S1/c1_S2}
  Q =  \frac{(c_1^a)^2 \theta_1 N_1 + (c_1^b)^2 (1-\theta_1) N_1}{(c_1^a)^2 \theta_2 N_2 + (c_1^b)^2 (1-\theta_2) N_2}.
\end{equation}
Substituting the ratio $N_1/N_2$ from \eqref{eq:N1/N2} into \eqref{eq:c1_S1/c1_S2} yields
\begin{equation*}
    Q = \frac{(c_1^a)^2 \theta_1 + (c_1^b)^2 (1-\theta_1)}{(c_1^a)^2 \theta_2  + (c_1^b)^2 (1-\theta_2) } \cdot \frac{c_0^a \theta_2 + c_0^b (1-\theta_2) }{c_0^a \theta_1 + c_0^b (1-\theta_1)}.
\end{equation*}
Upon rearranging, it can be shown that $Q>1$ if and only if $c_0^b (c_1^a)^2 > c_0^a (c_1^b)^2$. The claim of this corollary follows by evaluating the FS coefficients using \eqref{eq:c0n}--\eqref{eq:cmn}.
\end{IEEEproof}

\color{black}
\bibliographystyle{IEEEtran_DWPT}
\bibliography{myabrv,power,kekatos,DWPT}
\end{document}